\documentclass[aps,showpacs, prl,twocolumn,superscriptaddress]{revtex4}%
\usepackage{epsfig,dsfont,amssymb,amsmath,amsthm,amsfonts,amsbsy,mathrsfs}
\usepackage{graphicx}
\usepackage{amsmath}
\usepackage{amssymb}
\begin{document}
\title{Two semiconducting three-dimensional all-sp$^{2}$ carbon allotropes}
\author{Chaoyu He}
\affiliation{Institute for Quantum Engineering and Micro-Nano Energy
Technology, Xiangtan University, Xiangtan 411105, China}
\author{L. Z. Sun}
\email{lzsun@xtu.edu.cn} \affiliation{Department of Physics, Xiangtan University,
Xiangtan 411105, China}
\author{C. X. Zhang}
\affiliation{Institute for Quantum Engineering and Micro-Nano Energy
Technology, Xiangtan University, Xiangtan 411105, China}
\author{J. X. Zhong}
\email{zhong.xtu@gmail.com}\affiliation{Institute for Quantum
Engineering and Micro-Nano Energy Technology, Xiangtan University,
Xiangtan 411105, China}
\date{\today}
\pacs{64.60.My£¬64.70.K-£¬71.15.Mb, 71.20.Mq}
\begin{abstract}
Using first-principles method, we investigate the energetic
stability, dynamic stability and electronic properties of two
three-dimensional (3D) all-sp$^{2}$ carbon allotropes,
sp$^{2}$-diamond and cubic-graphite. The cubic-graphite was
predicted by Michael O'Keeffe in 1992 (Phys. Rev. Lett., 68, 15,
1992.) possessing space group of Pn-3m (224), whereas the
sp$^{2}$-diamond with the space group Fd-3m (227) same as that of
diamond has not been reported before. Our results indicate that
sp$^{2}$-diamond is more stable than previously proposed K4-carbon
and T-carbon, and cubic-graphite is even more stable than superhard
M-carbon, W-carbon and Z-carbon. The calculations on vibrational
properties show that both structures are dynamically stable. Interestingly,
both sp$^{2}$-diamond and cubic-graphite behave as semiconductors which are contrary
to previously proposed all-sp$^{2}$ carbon allotropes.
The sp$^{2}$-diamond is a semiconductor with a direct band gap of 1.66 eV, and cubic-graphite is an indirect semiconductor with band gap of 2.89 eV.
The very lower densities and entirely sp$^{2}$ configures of sp$^2$-diamond and cubic-graphite can be potentially applied in hydrogen-storage,
photocatalysts and molecular sieves.\\
\end{abstract}
\maketitle \indent The searching for low energy carbon allotropes
has been of great interest in the past few years. Many superhard
carbon phases with remarkable stability have been proposed, such as
the cage-based FCC136 \cite{FCC136}, fluffy T-carbon \cite{Tcarbon},
superhard M-carbon \cite{M1, M2}, bct-C4 \cite{b1, b2, b3, b4, b5},
W-carbon \cite{W}, Z-carbon \cite{z1, z2, z3}, S-carbon \cite{s1,
s2} and other novel carbon allotropes \cite{n1, n2, n3, n4, n5, n6,
n7}. All the above mentioned carbon crystals are sp$^3$ bonded and
most of them are considered as the potential products of cold
compressing graphite \cite{exp1}. To search for superhard materials,
some attempts have been made on designing three-dimensional (3D) all-sp$^2$
bonded carbon crystals motivated by the belief that shorter bonds make solid harder.
The 3D all-sp$^2$ bonded carbon systems having bond lengths smaller than those in
diamond are expected to be superhard materials or even harder than diamond. In view of
the graphite is intrinsically soft due to its layered configuration, some 3D network of
all-sp$^2$ bonded carbon atoms such as bct-4 \cite{4, 5}, H6 \cite{5, 6}, K4-carbon \cite{5, 7, CK4},
C-20 \cite{8} and cubic-graphite (6.8$^2$D) \cite{9} have been proposed. Although only the cubic-graphite
which is more stable than C60 has been successfully synthesized \cite{10, 11} in experiments, and none of
these all-sp$^2$ carbon phases have been announced harder than diamond, these novel all-sp$^2$ carbon networks
have given rise to many interests in material sciences \cite{f1, f2, f3, f4, f5}.\\
%**********************************************************************************************************************
\begin{figure}
\includegraphics[width=3.50in]{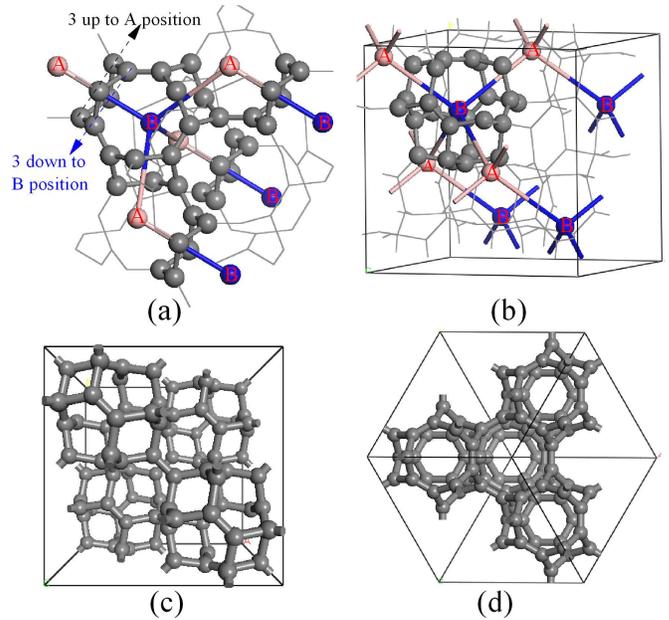}\\
\caption{Sketches in molecule form (a) and crystal form (b) of substituting each C-C bond in diamond with distorted C6 members ensuring every six carbon atoms in each C6 form a ``3up/3down'' configuration; The crystalline views of optimized sp$^2$-diamond from [001] direction (c) and [111] direction.}\label{fig1}
\end{figure}
%**********************************************************************************************************************
\indent In this paper, we propose a stable 3D all-sp$^{2}$ carbon allotrope named as sp$^{2}$-diamond whose space group (Fd-3m (227))
is the same as that of diamond. Our calculations reveal that sp$^{2}$-diamond is more stable than previously proposed T-carbon, H6, K4-carbon
and C20 but less stable than bct4 and the most stable cubic graphite. Vibrational properties of sp$^{2}$-diamond and cubic-graphite indicate their
dynamically stability. Interestingly, the results of the electronic structures show their semiconducting characteristics. The cubic graphite is an
indirect band-gap semiconductor with a gap of 2.891 eV and sp$^{2}$-diamond is a direct band gap semiconductor with gap of 1.66eV. However, all
previously proposed bct4, H6, K4-carbon and C-20 are metals.\\
%*************************figure 2*************************************************************************
\begin{figure}
\includegraphics[width=3.5in]{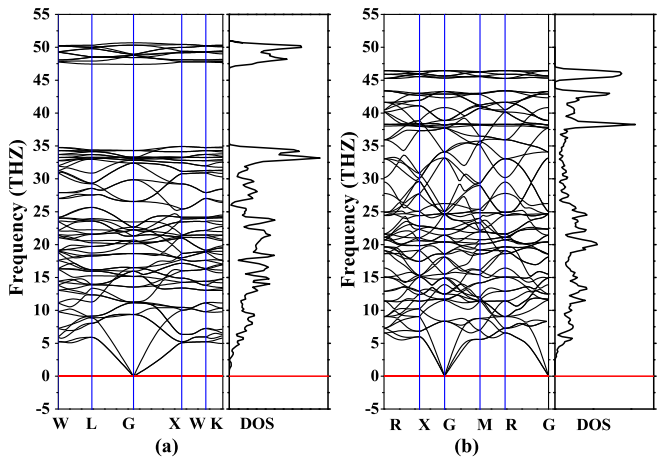}\\
\caption{Phonon band structures and phonon density
states of sp$^2$-diamond (a) and cubic-graphite(b).}\label{fig2}
\end{figure}
%**********************************************************************************************************************
\indent All calculations are performed using the density functional theory based VASP code \cite{vasp} with the projected
augmented wave (PAW) potential \cite{paw}. The exchange and correlation are approximated by general gradient approximation (GGA)
developed by Perdew et al. \cite{GGA}. The wave functions for all systems are expanded by plane-wave functions with cutoff energy
of 500 eV. The Brillouin zone sample meshes based on the Monkhorst-Pack scheme are set to be 9x9x9 for C20, cubic graphite and sp$^{2}$-diamond,
13x13x13 for K4-carbon and cubic-diamond and 13x13x5 for graphite, H6 and bct-4. Lattice constants and atom positions for all allotropes considered
in present work are fully optimized until the residual force on each atom less than 0.001 eV/A. The calculations of phonon band structures and phonon
density of states (DOS) are performed using the phonon \cite{phonon} package with applying forces from VASP calculations.\\
%**********************************************************************************************
%+++++++++++++table I ++++++++++++++++++++++++++++++++%
\begin{table*}
  \centering
  \caption{Space group, Lattice constant (LC: {\AA}), inequivalent positions (POS), bond length (L$_B$: {\AA}),
  mass density ($\rho$: g/cm$^{3}$), cohesive energy (Ecoh: eV) and band gap (Eg: eV) of graphite, diamond,
  T-carbon, bct4, H-6, K4-carbon, C-20, sp$^{2}$-diamond and cubic-graphite.}\label{tabI}
\begin{tabular}{c c c c c c c c}
\hline \hline
Systems      &Space group       &LC                    &POS           &L$_B$        &$\rho$ &Ecoh &E$_g$\\
\hline
Graphite        &P63/mmc  &a=b=2.468, c=6.913  &(1.000, 0.000, 0.750) &1.425         &1.796 &-7.825  &0\\
                &         &                    &(0.197, 0.197, 1.000) &              &      &        & \\
diamond         &Fd-3m    &a=b=c=3.574         &(0.250, 0.250, 0.250) &1.548         &3.491 &-7.668  &4.635\\
T-carbon        &Fd-3m    &a=b=c=7.517         &(0.321, 0.321, 0.679) &1.416, 1.501  &1.501 &-6.519  &2.253\\
bct-4           &I41/amd  &a=b=2.538, c=8.666  &(1.000, 1.000, 0.918) &1.424, 1.470  &2.853 &-7.236  &metal\\
H6              &P6222    &a=b=2.645, c=6.374  &(0.500, 0.500, 0.947) &1.454, 1.483  &3.093 &-6.906  &metal\\
K4-carbon       &I4132    &a=b=c=4.126         &(0.125, 0.125, 0.125) &1.459         &2.269 &-6.529  &metal\\
C-20            &Fm-3m    &a=b=c=9.145         &(0.139, 0.139, 0.861) &1.354, 1.481  &2.084 &-6.878  &metal\\
                &         &a=b=c=9.145         &(0.197, 0.197, 1.000) &              &      &        &     \\
sp$^{2}$-diamond&Fd-3m    &a=b=c=9.668         &(0.451, 0.451, 0.726) &1.347, 1.506  &2.116 &-7.179  &1.663\\
cubic-graphite  &Pn-3m   &a=b=c=6.095          &(0.500, 0.086, 0.586) &1.408, 1.493  &2.111 &-7.585  &2.891\\
\hline \hline
\end{tabular}
\end{table*}
%***********************************************************************
\indent The very fluffy T-Carbon was proposed by substituting each carbon atom in diamond with a carbon tetrahedron and keeping
the same space group Fd-3m as diamond. Inspired by such a block skill of substitution, we construct a entirely sp$^2$ bonded carbon
network through substituting each C-C bond in a hypothetically enlarged diamond with distorted C6 members in proper directions and positions
keeping the same space group Fd-3m as diamond, as shown in Fig.~\ref{fig1} (a) and (b). Along with the substitution, the six carbon atoms
of each C6 member symmetrically distributed along the original C-C bond forming a ``3up/3down'' bonding configuration as indicated
in Fig.~\ref{fig1} (a). After optimization, all the C6 members in this crystal are equivalent and connect to each other through
inter-C6 C-C bonds with length of 1.347 {\AA}. The lattice constant of sp$^{2}$-diamond is 9.668 {\AA} and its cohesive energy
is -7.179 eV/atom which is 660 meV/atom lower than that of T-carbon. In Fig.~\ref{fig1} (c) and (d) we show the views of optimized
sp$^2$-diamond from [001] direction and [111] direction, respectively. In the crystal cell of sp$^2$-diamond, there is only one equivalent
carbon atom locating at the position of (0.451, 0.451, 0.726). Carbon atoms in sp$^{2}$-diamond form equivalent C6 members through intra-C6
bonds with length of 1.506 {\AA}, and these equivalent C6 members connect to each other forming a periodic 3D all-sp$^{2}$ carbon network. Such
structural characteristics are very similar to those of cubic-graphite, in which equivalent carbon atoms form equivalent C6 members with intra-C6
bonds and equivalent C6 members connect to each other through inter-C6 bonds, forming pure sp$^2$ carbon network with space group of Pn-3m (224).
The lattice constant of cubic-graphite is 6.095 {\AA} and it contains only one inequivalent carbon atom in its crystal cell locating at the position
of (0.500, 0.086, 0.586). Its intra-C6 bond and inter-C6 bond lengths are 1.408 {\AA} and 1.493 {\AA}, respectively. The major difference between
sp$^2$-diamond and cubic-graphite is that in cubic-graphite the six carbon atoms in each C6 member are planar forming standard sp$^2$ hybridization,
whereas in sp$^2$-diamond the six carbon atoms in each C6 member  form a ``3up/3down'' configuration with distorted sp$^2$ bonds.\\
%********************************************************************
%*************************figure 3*************************************************************************
\begin{figure}
\includegraphics[width=3.50in]{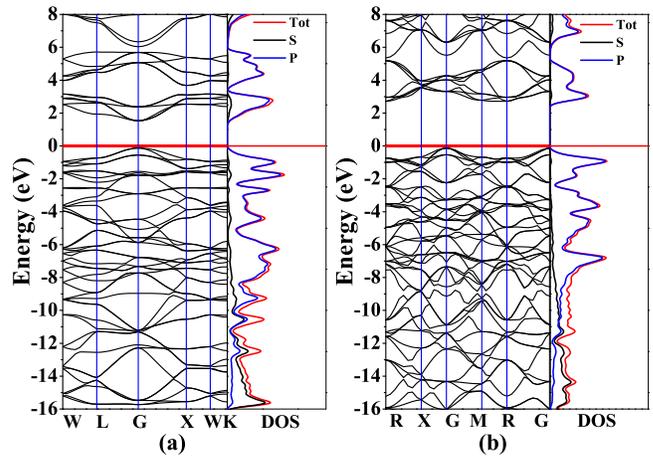}
\caption{Electronic band structures and density
states of sp$^2$-diamond (a) and cubic-graphite(b).}\label{fig3}
\end{figure}
\indent The cohesive energies, electronic properties (metal or semiconductor) and the structure information including space group,
lattice constants, atom positions, bond lengthes and mass density for sp$^{2}$-diamond and cubic-graphite are listed in Tab I. From
Tab I, we can see that the cohesive energies of graphite, diamond, T-carbon, bct4, H6, K4, C20, sp$^2$-diamond and cubic-graphite
are -7.825 eV/atom, -7.668 eV/atom, -6.519 eV/atom, -7.236 eV/atom, -6.906 eV/atom, -6.529 eV/atom, -6.878 eV/atom, -7.179 eV/atom
and -7.585 eV/atom, respectively. The cubic-graphite (-7.585 eV/atom) is more stable than the superhard M-carbon (-7.531 eV/atom),
W-carbon (-7.540 eV/atom) and H-carbon (-7.554 eV/atom). Although sp$^2$-diamond is less stable than bct4, cubic-graphite, diamond
and graphite, it is more stable than H6, K4, C20 and sp$^3$ bonded T-carbon. We then examine the dynamic stability of sp$^2$-diamond
and cubic-graphite through simulating their vibrational properties. The calculated phonon band structures and phonon density of states
are shown in Fig.~\ref{fig2} (a) and (b) for sp$^2$-diamond and cubic-graphite, respectively. We can see that there are no negative
modes in both sp$^2$-diamond and cubic-graphite, confirming both structures are dynamically stable. The very lower densities of
2.116 g/cm$^3$ and 2.111 g/cm$^3$ for sp$^{2}$-diamond and cubic-graphite as well as their porous configurations indicate that both
of them are sparse materials hoping to be applied for hydrogen-storage, catalysts and molecular sieves. \\
%*************************figure 2*************************************************************************
\begin{figure}
\includegraphics[width=3.50in]{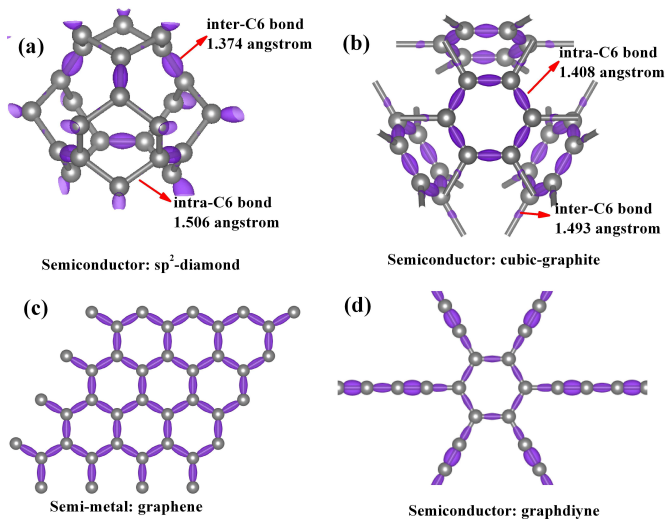}\\
\caption{Bonding charge density (isovalues = 0.007e/${\AA}^3$, ) of
sp$^2$-diamond (a), cubic-graphite (b), graphene (c) and graphdiyne
(d).}\label{fig4}
\end{figure}
%****************************************************************************************************
\indent Usually, quasi-1D carbon phases with only sp$^{2}$
hybridization (carbon nanotubes) can be metals or semiconductors
dependent on their helicities \cite{cnt1}. 2D graphene
allotropes\cite{2D1, 2D2, 2D3, 2D4} with only sp$^{2}$ bonds are
semi-metals or metals except for the semiconducting one proposed by
Mark et al \cite{2DSem}. 2D graphdiyne with both sp$^{2}$ and sp
hybridized bonds is semiconducting. Almost all the previously
proposed 3D carbon allotropes with pure sp$^2$ network are metals.
Interestingly, we find that both sp$^{2}$-diamond and cubic-graphite
with only sp$^2$ bonds are semiconductors. Electronic band
structures and density of states (DOS) of sp$^{2}$-diamond and
cubic-graphite are shown in Fig.~\ref{fig3} (a) and (b),
respectively. We can see that sp$^{2}$-diamond is a direct-band-gap
semiconductor with a gap of 1.66 eV and cubic-graphite is an
indirect-band-gap semiconductor with a larger gap of 2.89 eV. From
their projected DOS (PDOS), we find that the states around the
Fermi-level equally contributed from 2p$_x$, 2p$_y$ and 2p$_z$
states, and they are much larger than those derived from 2s orbital.
2s orbital electrons states mainly distribute at energy area about
12 eV below the Fermi-level. To understand the novel semiconducting
properties of sp$^{2}$-diamond and cubic-graphite, we investigate
the bonding charge density of these two systems through comparing
with semi-metallic graphene and semiconducting graphdiyne. The
bonding charge density is defined as the difference between the
total charge density in the solid and the superpositions of neutral
atomic charge densities placed at atomic sites, i.e.,
\begin{eqnarray}
\Delta \rho (r)=\rho _{solid}(r)-{\sum_\alpha}\rho _\alpha
(r-r_\alpha ) \label{eq:one}
\end{eqnarray}
Therefore, the bonding charge density represents the net charge redistribution as atoms are brought together to form the crystal.
Fig.~\ref{fig4} (a), (b), (c) and (d) show the bonding charge density of sp$^2$-diamond, cubic-graphite, graphene and graphdiyne, respectively.
We can see that the charge density uniformly distribute on the equivalent C-C bonds for semi-metallic graphene. For the semiconducting graphdiyne
with four inequivalent C-C bonds, electrons prefer locating at the shorter trinary C$\equiv$C bonds than other single C-C bonds, leading a semiconducting
property. In sp$^2$-diamond (cubic-graphite), the inter-C6 (intra-C6) C-C bonds with shorter length hold more electrons than the intra-C6 (inter-C6) C-C bonds.
The bonding characters of both sp$^2$-diamond and cubic-graphite are similar to those of graphdiyne, indicating that the un-bonded fourth-electrons of each
carbon atoms are not dissociative as those in semi-metallic graphene and graphite. So sp$^2$-diamond and cubic-graphite behave as semiconductors. \\
%**************************************************************************************
\begin{figure}
\includegraphics[width=3.50in]{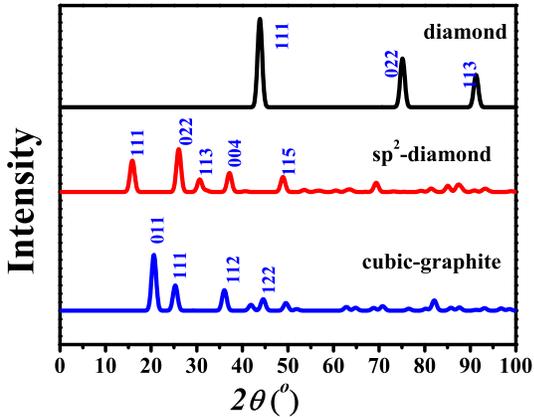}
\caption{Simulated x-ray diffraction patterns for diamond,
sp$^2$-diamond and cubic-graphite.}\label{fig5}
\end{figure}
%**************************************************************************************
\indent Both sp$^2$-diamond and cubic-graphite have cubic lattice as the same of diamond. To experimentally identify the two new forms of carbon, we provide the
simulated x-ray (with wavelength of 1.4059 {\AA}) diffraction (XRD) patterns for diamond, sp$^{2}$-diamond and cubic-graphite as shown in Fig.~\ref{fig4}. We can
see that the XRD patterns of sp$^2$-diamond and cubic-graphite can be easily distinguished from that of diamond. sp$^{2}$-diamond possess the same space group of
diamond (Fd-3m). In its XRD pattern, three peaks of (111), (022) and (113) mainly distribute at the area of 2$\theta$=$15^{o}$-$35^{o}$. Differently, these three
peaks appear in XRD pattern of diamond locating at 2$\theta$=$43.84^{o}$, $75.13^{o}$ and $91.26^{o}$, respectively. cubic-graphite belongs to space group Pn-3m (224)
and its XRD pattern contains peaks located at 2$\theta$=$20.59^{o}$ (011), $25.28^{o}$ (111), $36.04^{o}$ (112) and $44.56^{o}$ (122). These results are helpful for
identifying the sp$^{2}$-diamond and cubic-graphite in experiment.\\
%**************************************************************************************
\indent In summary, we proposed a 3D all-sp$^{2}$ carbon allotropes
(sp$^2$-diamond) with intriguing structure, remarkable stability.
The dynamical stability of sp$^2$-diamond and the previously
proposed cubic-graphite are confirmed by simulating their
vibrational properties. Both sp$^2$-diamond and cubic-graphite are
semiconductors with direct and indirect band gaps of 1.66 eV and
2.89 eV, respectively. Such semiconducting characteristics of
sp$^2$-diamond and cubic-graphite are contrary to previously
proposed all-sp$^{2}$ carbon allotropes and the intuitive
notion of the electronic characteristics of carbons with sp$^{2}$
bonding nature. The very lower densities of 2.116 g/cm$^3$ and
2.111 g/cm$^3$ for sp$^{2}$-diamond and cubic-graphite as well as
their porous configurations indicate that both of them are sparse
materials hoping to be applied in hydrogen-storage, catalysts and molecular sieves.\\
%**************************************************************************************
\begin{acknowledgments}
This work is supported by the National Natural Science Foundation of China (Grant Nos. 51172191, 10874143, and 11074211),  the Program for New Century Excellent Talents in University (Grant No. NCET-10-0169), and the Scientific Research
Fund of Hunan Provincial Education Department (Grant Nos. 10K065, 10A118) \\
\end{acknowledgments}
%+++++++++++++++++++++++++++++++++++++++++++++++++++++++++

\end{document}